\documentclass[submission,copyright,creativecommons]{eptcs}
\providecommand{\event}{REF 2015} 
\usepackage{breakurl}             

\usepackage{itma}
\usepackage{amstext}
\usepackage{amsmath}
\usepackage{amsopn}
\usepackage{amssymb}
\usepackage{amscd}
\usepackage{graphicx}
\usepackage{color}

\newcommand{\OR}{\mathbin{\mkern 1mu [\!] \mkern 1mu}}
\newcommand{\POR}{\;/\!\!/\;}
\newcommand{\is}{\mathbin{:\raisebox{0.10em}{${\scriptscriptstyle =}$}}}

\newcommand{\bnfeq}{\mathrel{{:}{:}{=}}}

\newcommand{\SEQ}{\mathbin{\mkern 1mu \sf ; \mkern 1mu}}
\newcommand{ \q}[1]{{\mathcal#1}}
\newcommand{\VAR}{{\sf var\ }}
\newcommand{\st}{\bullet}
\newcommand{\DO}{{\sf do\ }}
\newcommand{\OD}{{\sf od\ }}

\newcommand{\DEP}{{\backslash\!\!\backslash}}
\renewcommand{\implies}{\Rightarrow}
\newcommand{\REF}{\sqsubseteq}
\providecommand{\urlalt}[2]{\href{#1}{#2}}
\providecommand{\doi}[1]{doi:\urlalt{http://dx.doi.org/#1}{#1}}

\title{A Theory of Service Dependency}
\author{Mats Neovius \qquad\qquad Luigia Petre \qquad\qquad Kaisa Sere
\institute{\AA bo Akademi University, Faculty of Science and Engineering\\Turku Centre for Computer Science (TUCS)\\ Turku, Finland}
}
\def\titlerunning{A Theory of Service Dependency}
\def\authorrunning{M. Neovius, L. Petre \& K. Sere}
\begin{document}
\maketitle

\begin{abstract}
Service composition has become commonplace nowadays, in large part due to the increased complexity of software and supporting networks. Composition can be of many types, for instance sequential, prioritising, non-deterministic. However, a fundamental feature of the services to be composed consists in their dependencies with respect to each other. In this paper we propose a theory of service dependency, modelled around a dependency operator in the Action Systems formalism. We analyze its properties, composition behaviour, and refinement conditions with accompanying examples. \end{abstract}

\section{Introduction}
Dependency can be of several types, for instance we can think of type and format dependencies between data producers and data consumers or of signature and semantic dependencies between service providers and service users. Moreover, when getting serviced by various service providers, we depend on them in ways not yet formally understood. As our contemporary digital activities (such as online banking or shopping) are based on service providers, that use in their turn other service providers, we need to better understand the composition types between all the involved services. More interestingly, composing services that depend on each other in various ways adds a special flavor to the problem.

In this paper we define dependency via a specific operator and analyze its properties especially in correlation with previously defined and studied composition operators. Our study is developed in the Action Systems \cite{BaKu83} formal framework. Analysis includes examining basic algebraic properties in the formal Action Systems framework as well as detailing how refinement applies to dependency.

Action Systems is a state-based formal method for modeling distributed systems. Introduced in 1983, when CSP \cite{Hoa78} and CCS \cite{Mil80} where the major modeling formalisms, it differed from them in that it proposed an overall approach of a system. CSP and CCS are process algebras, modeling the behavior of the processes of a system, together with their interactions. The basic idea of Action Systems is to model the overall system behavior, often in an abstract manner. The genericity of such abstractions are not problematic because Action Systems is built around the concept of {\em refinement}: a specification can be correctly developed from a more abstract to a more concrete form, by following formal rules for such developments. Nowadays, Action Systems are very resemblant of the Event-B \cite{eventbbook} formal method, which is, in fact, based on it and on the B-method \cite{bbook}. Notable for Event-B is an associated theorem prover, the Rodin platform \cite{rodin}, in which one can edit the system models and get automatically the proof obligations to prove, in order for the models to be correct with respect to various properties. Action Systems remains to this day much more general and flexible than Event-B, even if it has the downside of missing an equivallent tool platform. However, we set up our study of dependency in Action Systems, because of its flexibility. Once we understand all the concepts well, we plan to move our understanding into the Rodin platform, in form of a theory of actions. This is still very preliminary; we mention some thoughts on this in the conclusions.

Hence, the contribution of this paper consists in modelling dependency in a state-based formalism via a dedicated operator. We analyse fundamental properties of this operator, including refinement, and emphasise various examples relevant to our discussion. We believe this is the beginning of a series of studies on dependency, as it has become such an intricate phenomenon in our digitalised society.

We proceed as follows. In Section 2 we discuss Action Systems to the extent needed in this paper, after which in Section 3 we introduce and analyse the dependency operator. Refinement laws for dependency are studied in Section 4, and in Section 5 we outline an example that illustrates some of the introduced concepts. We conclude the paper with highlights of future work in Section 6.

\section{Action Systems - a Revisit}\label{as-revisit}
The Action System framework was introduced by Back and Kurki-Suonio in  1983~\cite{BaKu83} for modeling distributed systems. It has been investigated and extensively developed for about two decades, prominently by Back, Sere, von Wright, Sekerinski, Butler, and colleagues~\cite{procBaSe,modules,mpc89,BSS,HKS97,rosese96,Sekerinski96,SS96,SeWa97,WaSe98}. An Action Systems overview appears in~\cite{LPthesis05}.

In the following we revisit some of the fundamental building blocks of Action Systems that will then be employed for studying dependency.

\subsection{Preliminaries}\label{as-prelim}
An {\em action system} consists of a \emph{state} that can be evaluated and modified by a finite set of {\em actions}. The state models the problem domain of the system via a finite set of \emph{variables}: at any moment, the state contains the values of these variables. The state can also be described as a predicate understood as the conjunction of predicates describing the values of the variables, for instance the state can be described as $x=5 \wedge y=10$, where $x$ and $y$ are the state variables. The value of a variable can be read and modified by an action. Each action can read and modify a subset of the state variables. An action system is not necessarily regarded in isolation, but as a part of a more complex system. The rest of the system (the {\em environment}) communicates with the action system using different mechanisms such as global variables or exported procedures~\cite{SeWa97}.

An action system $\q{A}$ has the following form:
\begin{eqnarray}
\label{as_synt} \mathcal{A}&=&|[\VAR x \st S_0 \SEQ \DO A_1\OR\ldots\OR A_m\
\OD]|\ y
\end{eqnarray}
Here $x=x_1,\ldots,x_n$ are the {\em variables} of the system $\q{A}$, $S_0$ is a statement initializing them, while $A_i$, $i=1,\ldots,m$, are the {\em actions} of the system. Variables in $x$ may be {\em exported}, in the sense that they can be read, or written, or both read and written by environment actions. We refer to a local variable of $\q{A}$ that is not exported as \emph{private}. The {\em imported variables} $y=y_1,\ldots,y_k$ are declared in the environment of $\q{A}$. We assume that $x \cap y =\emptyset$ and refer to $x \cup y$ as the \emph{global} variables of $\q{A}$.

Notation-wise, we observe that the boundary of the system $\mathcal{A}$ is denoted with brackets $|[ ... ]|$; the entities inside the brackets are defined within $\mathcal{A}$ while the entities outside the brackets (namely, $y$) are not. Inside $\mathcal{A}$ we observe the sequence $\SEQ$ with its two parts: the first consists of the variable declaration and initialisation and the second consists of a $\DO ...\ \OD$ loop containing the actions separated by the non-deterministic choice operator $\OR$. We explain the execution model shortly, upon understanding the form and semantics of actions.

An $action$ is an atomic statement that can change the values of the local or global variables of the action system. An action $A$ can be described by the following grammar:
\begin{eqnarray}
\label{actions}
A & \bnfeq & abort \mid skip \mid x \is v \mid p\!\to A  \mid A \OR A.
\end{eqnarray}
Here $x$ is a list of variables, $v$ a list of values, and $p$ a predicate on the state variables. Intuitively, `$abort$' is the action that always deadlocks, `$skip$' is the stuttering action, `$x \is v$' is a multiple assignment, `$p\!\to A$' is a guarded action, executable only when $p$ holds, and `$A_1 \OR A_2$' is the nondeterministic choice among actions `$A_1$' and `$A_2$'. We note here that the actions~\eqref{actions} are suitable for specification, being rather abstract. For instance, the more deterministic sequential and prioritising compositions are missing, although they are well known for Action Systems. We discuss them shortly.

The semantics of an action $A$ is described in terms of the weakest precondition predicate transformer $wp$, in the style of Dijkstra~\cite{Dijkstra76}. The weakest precondition predicate transformer relates the state of the system after an action $A$ has taken place (the postcondition $q$ of $A$) to the widest possible state of the system before the action $A$ has taken place (weakest precondition of $A$ with respect to $q$). In this way, it completely describes an action by defining from what precondition one should start in order to arrive at a desired postcondition. Given a postcondition $q$, the function $wp(A,q)$ is defined below for actions~\eqref{actions}:

\[
\begin{array}{@{}lcl@{}} wp(abort,q) &=& false\\
wp(skip,q) &=& q\\
wp(x \is v,q) &=&  q[x:=v]\\
wp(p\!\to A,q) &=&(p \Rightarrow wp(A,q))\\
wp(A_1 \OR A_2,q) &=& wp(A_1,q) \wedge wp(A_2,q).
\end{array}\]

\noindent The details of the definition of this function are studied elsewhere~\cite{procBaSe,SeWa97}. Here we just discuss an intuitive understanding of this semantical way of defining actions. Consider action $skip$: it is clear that, to arrive at a postcondition $q$ by doing nothing in terms of state changes, one should start from the same precondition $q$. Consider also the assignment $x \is v$: what we want to happen when such an assignment is executed is that the variables $x$ should end up with values $v$ and all the other variables should keep their values; hence, to arrive at a postcondition $q$ after executing $x \is v$, we need to replace all occurrences of $x$ in $q$ with $v$. The action $abort$ is a special case, denoting an abandonment of computation; to arrive at postcondition $q$ when such an abandonment takes place is impossible, hence, there is no state from where one could get to $q$ via $abort$; hence, $wp(abort,q)=false$. For action $p\!\to A$, we execute $A$ to get to $q$, but only when $p$ holds; hence, we need to start in a state where $wp(A,q)$ holds when $p$ holds. When $p$ does not hold, then nothing happens, so one can start from anything; in this case `anything' is modelled by $true$. The nondeterministic choice $A_1 \OR A_2$ is perhaps the most interesting of the actions~\eqref{actions}, as it means that either $A_1$ or $A_2$ can take place, but there is no way to know in advance which of them actually takes place; hence, in order to get to $q$ with $A_1 \OR A_2$ we must be prepared and start from a state in which both $wp(A_1,q)$ and $wp(A_2,q)$ hold.

A useful property of the $wp$ predicate transformer is to characterize the termination of an action: if $wp(A,true)=true$, then we say that $A$ terminates (also referred to as \emph{always terminating}~\cite{SeWa97}). This can be interpreted so that $true$ is a postcondition describing the state of the system without any restriction: the variables could have any values. Thus, if for an action $A$ we have that $wp(A,true)=true$, then we know nothing about how this action works except that it terminates.

The predicate transformer $wp$ is also conjunctive, as defined below; this property is useful when doing $wp$-calculations, as we will see in Section 3.
\begin{equation}
\label{conjunctive}
 wp(A, p \wedge q) = wp(A,p) \wedge wp(A,q)
\end{equation}

\paragraph{Equality of actions}
\label{comp}
Based on the $wp$ predicate transformer we can compare various (compositions of) actions. We are interested only in the input-output behaviour of actions, in terms of state changes, hence we consider two actions to be \emph{equal} if they always start from the same weakest precondition in order to achieve the same postcondition, for all possible postconditions:
\begin{equation}
\label{equality}
 A_1 = A_2 \ \ \ \ \ \ \text{iff} \ \ \ \ \ \ \text{for all}\   q : \ \ wp(A_1, q) = wp(A_2, q)
\end{equation}

\paragraph{Enabledness} An important property of an action is its {\em enabledness}, defined via the action's \emph{guard}: we say that an action is \emph{enabled} when its guard holds.
We are interested in `functional' states when modeling, namely those from where actions achieve useful postconditions; for this, we exclude those states from which an action establishes postcondition $false$, which models an impossible state. Hence, we define the guard of $A$, denoted $g(A)$, as $\neg wp(A,false)$: this gives those states in which action $A$ behaves in a functional way. The actions~\eqref{actions} have the following guards:
\begin{equation}
\label{guards}
\begin{array}{@{}lcl@{}}
g(abort) & = & true\\
g(skip)  & = & true\\
g(x \is v) & = & true\\
g(p\!\to A)  & = & p \wedge g(A)\\
g(A_1 \OR A_2) & = & g(A_1) \lor g(A_2)\\
\end{array}
\end{equation}
Actions whose guards are always true are called \emph{always enabled}~\cite{SeWa97}, for instance an assignment or a $skip$ action are always enabled. The action $p\!\to A$ is enabled when both $p$ and $g(A)$ hold: $\neg wp(p\!\to A,false)=\neg(p \Rightarrow wp(A,false))=\neg(\neg p \lor wp(A,false))=p \wedge \neg wp(A,false)=p \wedge g(A)$. A similar calculation leads to the formula~\eqref{guards} for $g(A_1 \OR A_2)$.

When we think of an action $A$ having the guard $g(A)$, the guardless `rest' of the action is syntactically referred to as the \emph{body} $b(A)$ of action $A$, so that $g(b(A))=true$. Thus, we can write action $A$ as $A=g(A)\to b(A)$. The study of action guards appears in the Action Systems literature, for instance in~\cite{procBaSe,Sekerinski96,SeWa97}, to support various other constructs. As enabledness is very important for service dependency, in this paper we study guards themselves in more detail. Notation-wise, whenever convenient we write $gA$ instead of $g(A)$ and similarly we write $bA$ instead of $b(A)$.

\begin{figure}[h]
\centering
\renewcommand{\figurename}{Figure}
\includegraphics[width=0.25\textwidth]{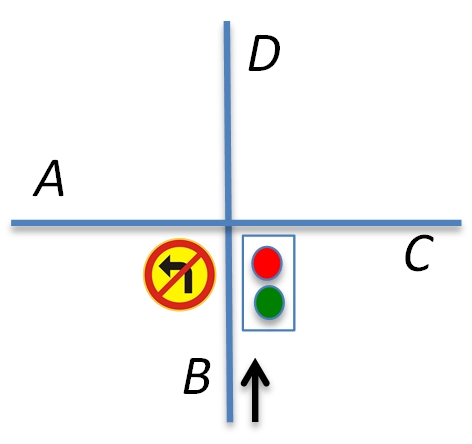}
\caption{A simple crossing with one car\label{CCB}}
\end{figure}

\paragraph{Example 1} Lets assume we have a simple road crossing as the one illustrated in Figure~\ref{CCB}. We model the two crossing roads as four segments labelled $A$, $B$, $C$, $D$. The action system $\mathcal{CC}{\!}{\!}_B$ below describes a simple behavior of a car entering the crossing at segment $B$.

\begin{eqnarray*}
\label{as_cc1}
\mathcal{CC}{\!}{\!}_B&=&|[\VAR light: \{green,\ red\}, loc:\{A,\ B,\ C,\ D\} \st \\
&& light,~loc \is red,~B  \SEQ \\
&&\DO \\
&&\ \ \!(A_1) \ light=red \to light \is green \\
&&\OR  (A_2) \ light=green \to light \is red \\
&&\OR  (A_3) \ loc=B \wedge light=green \to loc \is C \OR loc\is D \\
&&\OD]|
\end{eqnarray*}

\noindent The car on segment $B$ can only go through the crossing if the light is $green$. The state of $\mathcal{CC}{\!}{\!}_B$ is described by two variables $light$ and $loc$, the first modeling the crossing lights and the second the location of the car on one of the four segments $A$, $B$, $C$, $D$. We can see examples of assignments in this small Action System, as well as of guards and non-determinism.
Actions $A_1$ and $A_2$ switch between the lights. Action $A_3$ models that the location of the car can change from $B$ to $C$ or $D$ only when the car is at location $B$ ($loc=B$) and the crossing lights are green ($light=green$). In this case, the location is non-deterministically changed to either $C$ or $D$: $loc \is C \OR loc\is D$.

\subsection{The execution model}\label{as_exec_model}
The execution of an action system $\q{A}$ as in~\eqref{as_synt} is the following. The initiali\-sation $S_0$ sets the variables to some specific values. Then, from the enabled actions, one is non-deterministically chosen and executed: this means that the chosen action changes the values of its accessed variables in a way that is determined by the action body. The variables that are not accessed by that action keep their values unchanged. The execution of any action is atomic: this means that, once the action is selected for execution, it will execute without interference from other actions. The computation terminates when no action is enabled. This means that the state will evolve no more, fixing the final values of the variables forever. The action system $\mathcal{CC}{\!}{\!}_B$ is non-terminating, as the lights will keep switching via actions $A_1$ and $A_2$. Upon initialisation, after the lights become $green$, both actions $A_2$ and $A_3$ are enabled: if $A_2$ is chosen for execution, then the lights change back to $red$, and then only $A_1$ is enabled. When $lights=green$ and $A_3$ is chosen for execution, then the car will go straight on, advancing to segment $D$ or to the right, advancing to segment $C$.

Such an execution model is similar to Dijkstra's guarded iteration
statement~\cite{Dijkstra76}, showing Action Systems can model
\emph{sequential executions}. Parallelism can also be modelled in
the framework, by \emph{interleaving}. In such a \emph{parallel
execution} model, actions that do not access each other's variables
and are enabled at the same time can be executed in parallel. This
is possible because their sequential execution in any order has the
same result. Execution models are detailed in
\cite{parle89,rcII,mpc89}.

Execution of any action cannot be guaranteed in the Action System framework. This is due to assuming \emph{no notion of fairness} in the
model~\cite{parle89}. Fairness~\cite{MP83}, as a property that concurrent systems may have, can be of several forms. One of the most used forms,
referred also as \emph{strong} fairness, means that an action is infinitely often executed if it is infinitely often enabled. Having no
assumptions of fairness implies that true non-determinism can be modeled with Action Systems. Also, properties proved for a sequential execution
of an action system $\q{A}$ as in~\eqref{as_synt} still hold when a parallel execution is assumed for $\q{A}$~\cite{parle89,BaKu88}. This
feature is important because the theory supporting proofs about sequential executions is rich, see for example~\cite{Dijkstra76,Hoare69}.

\subsection{More deterministic composition operators}\label{as-detcomp}
The actions~\eqref{actions} can model abstract specifications, that include non-determinism as an abstraction mechanism. We now focus on two action composition operators that enable more determinism in the specifications. We extend the grammar~\eqref{actions} as follows:
\begin{eqnarray}
\label{actions1}
A & \bnfeq & \ldots \mid A \POR A \mid A \SEQ A .
\end{eqnarray}

\noindent Here `$A_1 \POR A_2$' is the prioritising composition of two actions `$A_1$' and `$A_2$' and `$A_1 \SEQ A_2$' is the sequential composition of two actions `$A_1$' and `$A_2$':
\begin{equation}
\label{detcompdef}
\begin{array}{@{}lcl@{}}
A_1 \POR A_2 & = & A_1 \OR \neg gA_1 \to A_2\\
A_1 \SEQ A_2 & = & gA_1 \wedge wp(bA_1,gA_2) \to bA_1 \SEQ bA_2
\end{array}
\end{equation}

\paragraph{Prioritising composition}
One reason behind defining the $\POR$ operator between actions if that of \emph{coordination}. The underlying execution model of Action Systems is non-deterministic, i.e., the
scheduling of certain actions for execution cannot be guaranteed. However, when modeling coordination we need to enforce the execution of specific actions. The notion of coordination was therefore defined in terms of prioritising composition for Action Systems in~\cite{HKS97,SS96}. We say that the action $A_1\ coordinates$ the action $A_2$. Essentially, action $A_1$ has a higher priority than action $A_2$: $A_1$ can be executed if it is enabled, while $A_2$ can be executed if it is enabled \emph{and} $A_1$ is not enabled.

\paragraph{Example 2} To see an example of prioritising composition, lets think again about a simple crossing, but this time without the crossing lights and with two cars trying to pass through it, as illustrated in Figure~\ref{CCBC}. Assume both cars just want to continue on their roads and so, without crossing lights, we need to take into account the right-of-way priority: the car coming from the East will have priority over the car coming from the South. We have the following two actions modeling the desired movement of the two cars:

\begin{figure}[h]
\centering
\renewcommand{\figurename}{Figure}
\includegraphics[width=0.25\textwidth]{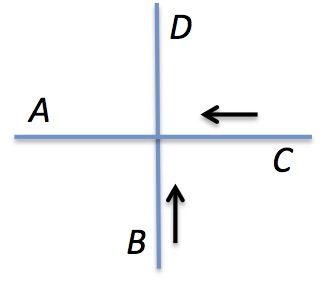}
\caption{A simple crossing with two cars\label{CCBC}}
\end{figure}

\begin{eqnarray*}
\label{act_cc2}
&(A_4) & loc_1=B \to loc_1 \is D\\
&(A_5) & loc_2=C \to loc_2 \is A
\end{eqnarray*}
Here, action $A_4$ models the desired movement of the car from the South, while action $A_5$ models the desired movement of the car from the East. Assuming right-hand traffic, the right-of-way is modeled by the prioritised composition $A_5 \POR A_4$: action $A_4$ will execute when enabled and when action $A_5$ is disabled.

The weakest precondition with respect to a predicate $q$ and the guard of $A_1 \POR A_2$ are, respectively:
\begin{equation}
\label{por}
\begin{array}{@{}lcl@{}}
wp(A_1 \POR A_2,q) &=& wp(A_1,q) \wedge (gA_1 \lor wp(A_2,q)) \\
g(A_1 \POR A_2) & = & gA_1 \lor gA_2\\
\end{array}
\end{equation}

\paragraph{Sequential composition}
For specifying sequentiality, we use the $\SEQ$ operator. This is, in fact, one of the fundamental operators in~\cite{Dijkstra76}, defined as follows: $A_1 \SEQ A_2$ behaves as $A_1$ if $A_1$ is enabled, then, when $A_1$ terminates, as $A_2$ if $A_2$ is enabled; otherwise, the sequence $A_1 \SEQ A_2$ is not enabled~\cite{BSS}. The weakest precondition with respect to a predicate $q$ and the guard of $A_1 \SEQ A_2$ are, respectively:
\begin{equation}
\label{seq}
\begin{array}{@{}lcl@{}}
wp(A_1 \SEQ A_2, q) &=& wp(A_1, wp(A_2, q)) \\
g(A_1 \SEQ A_2)  & = & gA_1 \wedge wp(bA_1,gA_2)\\
\end{array}
\end{equation}

A useful construct for working with actions is also the \emph{assumption} $[p]$, where $p$ is a predicate. We have that $wp([p], q) = (p \implies q)$ and that $g([p])=p$. The meaning of $[p]$ is that is behaves as $skip$ if $p$ holds and as $abort$ otherwise. Its usefulness comes from the fact that an action $p\!\to A$ is defined as $[p]\SEQ A$:
\begin{equation}
\label{assumpt}
\begin{array}{@{}lcl@{}}
p\!\to A &=& [p]\SEQ A
\end{array}
\end{equation}

Based on assumption and sequential composition, we define the following:
\begin{eqnarray}
\label{canenable} A_1\ \text{enables}\ A_2 &=& wp(A_1, gA_2) \\
\label{cannotdisable} A_1\ \text{cannot disable}\ A_2 &=& gA_2 \implies wp(A_1, gA_2) \\
\label{cannotenable} A_1\ \text{cannot enable}\ A_2 &=& \neg gA_2 \implies wp(A_1, \neg gA_2)
\end{eqnarray}
Definition~\eqref{canenable} essentially means that action $A_1$ terminates and establishes as postcondition $gA_2$: if $A_2$ was enabled before $A_1$ took place, then $A_1$ did not disabled it and if $A_2$ was disabled before $A_1$ took place, then $A_1$, via its state changes, enabled $A_2$. Understanding this feature is important for the dependency operator that we define in the next section. Essentially, $A_1\ \text{enables}\ A_2$ means that, if $A_1$ is enabled, then it will execute and enable $A_2$. This is also observable from the following calculation: $wp(A_1,gA_2)=wp([gA_1]\SEQ bA_1,gA_2)=wp([gA_1], wp(bA_1,gA_2))=(gA_1 \implies wp(bA_1,gA_2))$. Definition~\eqref{cannotdisable} is a stronger condition than~\eqref{canenable}, modelling that $A_2$ was enabled before $A_1$ took place and $A_1$ did not disabled it. More detailed calculations reduce definition~\eqref{cannotdisable} to $gA_1 \wedge gA_2 \implies wp(bA_1, gA_2)$. Definition~\eqref{cannotenable} models that, if $A_2$ was disabled before $A_1$ took place, then $A_1$ did not enabled it. More detailed calculations reduce definition~\eqref{cannotenable} to $gA_1 \wedge \neg gA_2 \implies wp(bA_1, \neg gA_2)$.

In the context of the above definitions, we rephrase the guard $g(A_1 \SEQ A_2)$ of $A_1 \SEQ A_2$~\eqref{seq} as follows: $A_1$ should be enabled and should enable $A_2$ upon its termination. We observe that $A_2$ does not need to be enabled before $A_1$ terminates.

We present below two more properties of the assumption construct, that are instrumental in this paper. First, we can split an assumption made of a conjunction of predicates into sequential (and commutative) assumptions, as described in~\eqref{split}:
\begin{equation}
\label{split}
\begin{array}{l}
\ \ wp([a \wedge b],q) \\
= \color{magenta}\{ \text{weakest precondition of assumption}\}\\
\ \ a \wedge b \implies q\\
= \color{magenta}\{ \text{logic}\}\\
\ \ \neg(a \wedge b) \vee q\\
= \color{magenta}\{ \text{logic}\}\\
\ \ (\neg a \vee \neg b) \vee q\\
= \color{magenta}\{\vee~\text{is associative}\}\\
\ \ \neg a \vee (\neg b \vee q)\\
= \color{magenta}\{ \text{logic}\}\\
\ \ a \implies (b \implies q)\\
= \color{magenta}\{ \text{weakest precondition of assumption, twice}\}\\
\ \ wp([a], wp([b],q)) \\
= \color{magenta}\{ \text{weakest precondition of}~\SEQ~\text{twice}\}\\
\ \ wp([a]\SEQ[b],q))
\end{array}
\end{equation}
\noindent Property~\eqref{split} can be thus written as $[a \wedge b] = [a] \SEQ [b]$ and this can also be written as $[a \wedge b] = [b] \SEQ [a]$, as conjunction and disjunction are commutative operators.

Second, we need to discuss what assumption $[wp(A,g)]$ means in a sequential composition. First, assume $wp(A,g)$ holds. This is then equivallent to $[true] \SEQ A \SEQ [g]$, meaning that, when $A$ is executed, it establishes postcondition $g$. Assume $wp(A,g)$ holds in the following sequential composition: $[wp(A,g)] \SEQ X \SEQ A \SEQ Y$. We can then rewrite $[wp(A,g)] \SEQ X \SEQ A \SEQ Y$ as $X \SEQ A \SEQ [g] \SEQ Y$. We formalize this in property~\eqref{key}:
\begin{equation}
\label{key}
\begin{array}{@{}lcl@{}}
[wp(A,g)] \SEQ X \SEQ A \SEQ Y &=& X \SEQ A \SEQ [g] \SEQ Y
\end{array}
\end{equation}

\section{The Dependency Operator}\label{dep-operator}
Having briefly reviewed the composition operators of actions, we now turn our attention to modelling dependency. The \emph{dependency} operator, $\DEP$, has already been introduced~\cite{NeSe08} as follows:
\begin{eqnarray}
\label{actions2}
A & \bnfeq & \ldots \mid A \DEP A,
\end{eqnarray}
\noindent where
\begin{equation}
\label{depdef}
\begin{array}{@{}lcl@{}}
A_1 \DEP A_2 & = & gA_1 \wedge gA_2 \to A_1 \SEQ A_2
\end{array}
\end{equation}
The weakest precondition with respect to a predicate $q$ and the guard of $A_1 \DEP A_2$ are, respectively:
\begin{equation}
\label{dep}
\begin{array}{@{}lcl@{}}
wp(A_1 \DEP A_2,q) &=& gA_1 \wedge gA_2 \implies wp(A_1, wp(A_2,q)) \\
g(A_1 \DEP A_2) & = & gA_1 \wedge gA_2 \wedge wp(bA_1,gA_2)\\
\end{array}
\end{equation}

The interpretation of this operator is as follows. We say that $A_1$ depends on $A_2$, because it needs $A_2$ to be enabled before and after its own ($A_1$'s) execution. In its turn, in the simple sequential composition $A_1 \SEQ A_2$ it is only necessary for $A_2$ to be enabled after $A_1$'s execution. To better see this difference between the two action compositions, we decompose $A_1 \SEQ A_2$ and $A_1 \DEP A_2$, based on ~\eqref{assumpt}, as follows:
\begin{eqnarray}
\label{diff} A_1 \SEQ A_2 &=& [gA_1] \SEQ bA_1 \SEQ [gA_2] \SEQ bA_2\\
\label{diff1} A_1 \DEP A_2 &=& [gA_1] \SEQ [gA_2] \SEQ bA_1 \SEQ [gA_2] \SEQ bA_2
\end{eqnarray}
Hence, $gA_2$ acts as an \emph{invariant} for $bA_1$, as observed in~\ref{diff1}: it should hold before and after $bA_1$ takes place.

\paragraph{Example 3} Let us assume the situation of a customer waiting to be served at a bank. In Finland, customers get each a queue number from a machine and wait for their number to be displayed on a screen, with an indication to which cashier to proceed. Once in front of the right cashiers, customers get serviced and get each a receipt for their service, printed by a printer at the cashier's desk. Assume we have three actions for a customer-server provider pair:
\begin{itemize}
  \item Action $A_1=gA_1 \to bA_1$, where $gA_1$ models that a customer has a queue number and $bA_1$ models that the customer waits to be served;
  \item Action $A_2=gA_2 \to bA_3$, where $gA_2$ models that the customer's number is called by a cashier (displayed on a screen) and $bA_2$ models that the cashier provides the desired service as well as commands the printing of the receipt;
  \item Action $A_3=gA_3 \to bA_3$, where $gA_3$ models that the (service provider's) printer has paper and $bA_3$ models that this printer prints the receipt.
\end{itemize}
We have obviously a sequence between $A_1$ and $A_2$: $A_1 \SEQ A_2 = [gA_1] \SEQ bA_1 \SEQ [gA_2] \SEQ bA_2$. The condition $gA_2$ (of the customer's number being called) does not need to hold before $bA_1$ takes place: it needs to hold after $bA_1$ took place. However, there is a dependency between $A_2$ and $A_3$: $A_2 \DEP A_3 = [gA_2] \SEQ [gA_3] \SEQ bA_2 \SEQ [gA_3] \SEQ bA_3$. The condition $gA_3$ that the printer has paper needs to hold before the cashier presses the `print' button on his screen; if $gA_3$ does not hold before that, then the cashier needs to do some other actions, for instance replenishing the paper.

We will now study several properties of the dependency operator. We begin by observing the following:
\begin{equation}
\label{dep_def2}
\begin{array}{l}
\ \ A_1 \DEP A_2 \\
= \color{magenta}\{ \text{definition}~\eqref{depdef}\ \text{of}\ \DEP\}\\
\ \ gA_1 \wedge gA_2 \to A_1 \SEQ A_2\\
= \color{magenta}\{\text{definition}~\eqref{detcompdef}\ \text{of}\ \SEQ\}\\
\ \ gA_1 \wedge gA_2 \to gA_1 \wedge wp(bA_1,gA_2) \to bA_1 \SEQ bA_2\\
= \color{magenta}\{\text{assumption definition}~\eqref{assumpt}  \}\\
\ \ [gA_1 \wedge gA_2] \SEQ [gA_1 \wedge wp(bA_1,gA_2)] \SEQ bA_1 \SEQ bA_2\\
= \color{magenta}\{\text{assumption properties, `$\SEQ$' is associative, logic}\}\\
\ \ [gA_1 \wedge gA_2 \wedge wp(bA_1,gA_2)]\SEQ bA_1 \SEQ bA_2\\
= \color{magenta}\{\text{assumption definition}~\eqref{assumpt}\}\\
\ \ gA_1 \wedge gA_2 \wedge wp(bA_1,gA_2) \to bA_1 \SEQ bA_2
\end{array}
\end{equation}
\noindent Hence, we can use either~\eqref{depdef},~\eqref{diff1} or~\eqref{dep_def2} to express $A_1 \DEP A_2$.

\paragraph{Commutativity and Associativity}
The dependency operator $\DEP$ is based on the sequential composition $\SEQ$. We know from basic theory that $A_1 \SEQ A_2 = A_2 \SEQ A_1$ only in special cases. In general, $\SEQ$ is not commutative and correspondingly, $\DEP$ is not commutative either.

We now consider associativity of $\DEP$. We have the following:
\begin{equation*}
\begin{array}{l}
\ \ (A_1 \DEP A_2) \DEP A_3 \\
= \color{magenta}\{\eqref{diff1}\}\\
\ \ ([gA_1] \SEQ [gA_2] \SEQ bA_1 \SEQ [gA_2] \SEQ bA_2)\DEP A_3\\
= \color{magenta}\{\eqref{diff1},\eqref{dep_def2}, \eqref{split}\}\\
\ \ [gA_1] \SEQ [gA_2] \SEQ [wp(bA_1,gA_2)] \SEQ [gA_3] \SEQ [wp(bA_1 \SEQ bA_2,gA_3)] \SEQ bA_1 \SEQ bA_2 \SEQ bA_3\\
= \color{magenta}\{\eqref{key}\}\\
\ \ [gA_1] \SEQ [gA_2] \SEQ [gA_3] \SEQ bA_1 \SEQ [gA_2] \SEQ bA_2 \SEQ [gA_3] \SEQ bA_3
\end{array}
\end{equation*}
By a similar computation, we have:
\begin{equation*}
\begin{array}{l}
\ \ A_1 \DEP (A_2 \DEP A_3)  \\
= \color{magenta}\{\eqref{diff1},\eqref{dep_def2}, \eqref{split},\eqref{key}\}\\
\ \ [gA_1] \SEQ [gA_2] \SEQ [gA_3] \SEQ bA_1 \SEQ [gA_2] {\bf \SEQ [gA_3]} \SEQ bA_2 \SEQ [gA_3] \SEQ bA_3
\end{array}
\end{equation*}
Since $\SEQ$ is associative, we obtain that $\DEP$ is associative only when $gA_3$ is an invariant for $A_1$, i.e., when $A_1$ cannot disable $A_3$~\eqref{cannotdisable}:
\begin{equation}
\label{dep_assoc}
{\bf (A_1 \DEP A_2) \DEP A_3 = A_1 \DEP (A_2 \DEP A_3) \ \ \ \ \ \ \text{iff} \ \ \ \ \ \ gA_3 \implies wp(A_1, gA_3)}
\end{equation}

\paragraph{Distributivity}

We now check the distributivity of $\DEP$ over $\SEQ$. We have the following:
\begin{equation*}
\begin{array}{l}
\ \ A_1 \DEP (A_2 \SEQ A_3) \\
= \color{magenta}\{\eqref{diff}\}\\
\ \ A_1 \DEP ([gA_2] \SEQ bA_2 \SEQ [gA_3] \SEQ bA_3)\\
= \color{magenta}\{\eqref{diff1}\}\\
\ \ [gA_1] \SEQ [gA_2] \SEQ bA_1 \SEQ [gA_2] \SEQ ([gA_2] \SEQ bA_2 \SEQ [gA_3] \SEQ bA_3)\\
= \color{magenta}\{\text{`$\SEQ$' is associative, logic}\}\\
\ \ ([gA_1] \SEQ [gA_2] \SEQ bA_1 \SEQ [gA_2] \SEQ bA_2) \SEQ [gA_3] \SEQ bA_3\\
= \color{magenta}\{\eqref{diff},\eqref{diff1}\}\\
\ \ (A_1 \DEP A_2) \SEQ A_3 \\
\end{array}
\end{equation*}
Hence, we have that:
\begin{equation}
\label{dep_seq_assoc}
{\bf A_1 \DEP (A_2 \SEQ A_3) = (A_1 \DEP A_2) \SEQ A_3}
\end{equation}
With respect to the distributivity of $\SEQ$ over $\DEP$, we have the following calculations:
\begin{equation*}
\begin{array}{l}
\ \ A_1 \SEQ (A_2 \DEP A_3) \\
= \color{magenta}\{\eqref{diff1}, \eqref{assumpt}\}\\
\ \ [gA_1] \SEQ bA_1 \SEQ ([gA_2] \SEQ [gA_3] \SEQ bA_2 \SEQ [gA_3] \SEQ bA_3)\\
= \color{magenta}\{\text{`$\SEQ$' is associative}, \eqref{split}\}\\
\ \ [gA_1] \SEQ {\bf bA_1 \SEQ [gA_3]} \SEQ [gA_2] \SEQ bA_2 \SEQ [gA_3] \SEQ bA_3
\end{array}
\end{equation*}
\begin{equation*}
\begin{array}{l}
\ \ (A_1 \SEQ A_2) \DEP A_3 \\
= \color{magenta}\{\eqref{diff}, \eqref{assumpt}\}\\
\ \ ([gA_1] \SEQ bA_1 \SEQ [gA_2] \SEQ bA_2) \DEP ([gA_3] \SEQ bA_3)\\
= \color{magenta}\{\text{`$\SEQ$' is associative}, \eqref{diff1}, \eqref{key}\}\\
\ \ [gA_1] \SEQ {\bf [gA_3] \SEQ bA_1} \SEQ [gA_2] \SEQ bA_2 \SEQ [gA_3] \SEQ bA_3
\end{array}
\end{equation*}
Hence, the associativity of $\SEQ$ over $\DEP$ holds only if $gA_3$ and $bA_1$ commute, for instance when $gA_3=true$ or when $bA_1$ and $gA_3$ have no variables in common. We thus have:
\begin{equation}
\label{seq_dep_assoc}
{\bf A_1 \SEQ (A_2 \DEP A_3) = (A_1 \SEQ A_2) \DEP A_3 \ \ \ \ \ \ \text{iff} \ \ \ \ \ \ [gA_3] \SEQ bA_1 = bA_1 \SEQ [gA_3]}
\end{equation}

We now check the distributivity of $\DEP$ over $\OR$ and $\POR$. We have the following:
\begin{equation*}
\begin{array}{l}
\ \ A_1 \DEP (A_2 \OR A_3) \\
= \color{magenta}\{\eqref{assumpt},\eqref{guards}\}\\
\ \ ([gA_1] \SEQ bA_1) \DEP ([gA_2 \lor gA_3] \SEQ (A_2 \OR A_3))\\
= \color{magenta}\{\eqref{diff1}\}\\
\ \ [gA_1] \SEQ [gA_2 \lor gA_3] \SEQ  bA_1 \SEQ [gA_2 \lor gA_3] \SEQ (A_2 \OR A_3)
\end{array}
\end{equation*}
\begin{equation*}
\begin{array}{l}
\ \ (A_1 \DEP A_2) \OR (A_1 \DEP A_3) \\
= \color{magenta}\{\eqref{diff1}\}\\
\ \ ([gA_1] \SEQ [gA_2] \SEQ bA_1 \SEQ [gA_2] \SEQ bA_2) \OR ([gA_1] \SEQ [gA_3] \SEQ bA_1 \SEQ [gA_3] \SEQ bA_3)\\
= \color{magenta}\{\eqref{assumpt}, \text{`$\SEQ$' is associative}, \eqref{split} \}\\
\ \ (gA_1 \wedge gA_2 \to bA_1 \SEQ A_2) \OR (gA_1 \wedge gA_3 \to bA_1 \SEQ A_3)\\
= \color{magenta}\{\text{properties of $\to$} \}\\
\ \ gA_1 \to ((gA_2 \to bA_1 \SEQ A_2) \OR (gA_3 \to bA_1 \SEQ A_3))\\
= \color{magenta}\{\eqref{assumpt}, \eqref{split}\}\\
\ \ [gA_1] \SEQ [gA_2 \lor gA_3] \SEQ bA_1 \SEQ [gA_2 \lor gA_3] \SEQ (A_2 \OR A_3)\\
\end{array}
\end{equation*}
Hence, $\DEP$ distributes over $\OR$ to the left:
\begin{equation}
\label{dep_par_left_distrib}
{\bf A_1 \DEP (A_2 \OR A_3) = (A_1 \DEP A_2) \OR (A_1 \DEP A_3)}
\end{equation}

By a similar proof we can show that $\DEP$ distributes over $\OR$ to the right as well. We do not show here the proof, for space purposes. Hence, we also have:
\begin{equation}
\label{dep_par_right_distrib}
{\bf (A_1 \OR A_2) \DEP A_3 = (A_1 \DEP A_3) \OR (A_2 \DEP A_3)}
\end{equation}

For checking the distributivity of $\DEP$ over $\POR$, we have the following calculations:
\begin{equation*}
\begin{array}{l}
\ \ A_1 \DEP (A_2 \POR A_3) \\
= \color{magenta}\{\eqref{detcompdef}\}\\
\ \ A_1 \DEP (A_2 \OR \neg gA_2 \to A_3)\\
= \color{magenta}\{\eqref{dep_par_left_distrib}\}\\
\ \ (A_1 \DEP A_2) \OR (A_1 \DEP (\neg gA_2 \to A_3))\\
= \color{magenta}\{\eqref{depdef}\}\\
\ \ (gA_2 \to A_1 \DEP A_2) \OR (\neg gA_2 \to A_1 \DEP \neg gA_2 \to A_3)\\
= \color{magenta}\{\eqref{detcompdef}\}\\
\ \ (A_1 \DEP A_2) \POR (A_1 \DEP \neg gA_2 \to A_3)
\end{array}
\end{equation*}
Hence, $\DEP$ distributes over $\POR$ to the left conditionally, if $A_1$ cannot enable $A_2$~\eqref{cannotenable}:
\begin{equation}
\label{dep_por_left_distrib}
{\bf A_1 \DEP (A_2 \POR A_3) = (A_1 \DEP A_2) \POR (A_1 \DEP A_3) \ \ \ \ \ \ \text{iff} \ \ \ \ \ \ \neg gA_2 \implies wp(A_1, \neg gA_2)}
\end{equation}
For distribution to the right we have:
\begin{equation*}
\begin{array}{l}
\ \ (A_1 \POR A_2) \DEP A_3 \\
= \color{magenta}\{\eqref{detcompdef}\}\\
\ \ (A_1 \OR \neg gA_1 \to A_2) \DEP A_3\\
= \color{magenta}\{\eqref{dep_par_right_distrib}\}\\
\ \ (A_1 \DEP A_3) \OR ((\neg gA_1 \to A_2) \DEP A_3)\\
= \color{magenta}\{\eqref{depdef}\}\\
\ \ (gA_1 \to A_1 \DEP A_3) \OR (\neg gA_1 \to A_2 \DEP A_3)\\
= \color{magenta}\{\eqref{detcompdef}\}\\
\ \ (A_1 \DEP A_3) \POR (A_2 \DEP A_3)
\end{array}
\end{equation*}
Hence, $\DEP$ distributes over $\POR$ to the right:
\begin{equation}
\label{dep_por_right_distrib}
{\bf (A_1 \POR A_2) \DEP A_3 = (A_1 \DEP A_3) \POR (A_2 \DEP A_3)}
\end{equation}


\section{Refinement}
We now shortly present the refinement relation $\REF$ between actions, discuss the relationship between $\SEQ$ and $\DEP$ with respect to refinement, as well as some monotonicity properties that are relevant for the dependency operator $\DEP$.

\paragraph{Refinement between actions} We say that action $A_1$ is \emph{refined} by action $A_2$, denoted $A_1 \REF A_2$, if the weakest precondition of the former implies the weakest precondition of the latter, with respect to the same postcondition $q$, for all postconditions $q$:
\begin{equation}
\label{refinement}
 A_1 \REF A_2 \ \ \ \ \ \ \text{iff} \ \ \ \ \ \ \text{for all}\   q : \ \ wp(A_1, q) \implies wp(A_2, q)
\end{equation}
The desired meaning of action refinement is that action $A_2$ is more deterministic than action $A_1$; this can be expressed as strengthening the guard ($gA_2 \implies gA_1$) and reducing non-determinism (e.g., $A_1 \OR A_2 \REF A_1 \POR A_2$).

\paragraph{Sequence is refined by dependency} We have that $A_1 \SEQ A_2 \REF A_1 \DEP A_2$, based on the following calculations:
\begin{equation*}
\begin{array}{l}
\ \ wp(A_1 \SEQ A_2, q)\\
= \color{magenta}\{\eqref{diff}\}\\
\ \ wp([gA_1] \SEQ bA_1 \SEQ [gA_2] \SEQ bA_2, q)\\
= \color{magenta}\{\eqref{seq},~\eqref{assumpt}\}\\
\ \ gA_1 \implies wp(bA_1, gA_2 \implies wp(bA_2, q))
\end{array}
\end{equation*}
\begin{equation*}
\begin{array}{l}
\ \ wp(A_1 \DEP A_2, q)\\
= \color{magenta}\{\eqref{diff1}\}\\
\ \ wp([gA_1] \SEQ [gA_2] \SEQ bA_1 \SEQ [gA_2] \SEQ bA_2, q)\\
= \color{magenta}\{\eqref{dep},~\eqref{assumpt}\}\\
\ \ gA_1 \wedge gA_2 \implies wp(bA_1, gA_2 \implies wp(bA_2, q))
\end{array}
\end{equation*}
If we denote $gA_1$ by $a$, $gA_2$ by $b$, and $wp(bA_1, gA_2 \implies wp(bA_2, q))$ by $c$, we have to show that $(a \implies c) \implies (a \wedge b \implies c)$, which holds. Hence, we have that:
\begin{equation}
\label{seqrefdep}
{\bf A_1 \SEQ A_2 \REF A_1 \DEP A_2}
\end{equation}

The reverse relation holds only if $gA_1 \implies gA_2$, hence:
\begin{equation}
\label{deprefseqcond}
{\bf A_1 \DEP A_2 \REF A_1 \SEQ A_2 \ \ \ \ \ \ \text{iff} \ \ \ \ \ \ gA_1 \implies gA_2}
\end{equation}

\paragraph{Monotonicity} Nondeterministic choice $\OR$ and sequential composition $\SEQ$ are monotonic with respect to refinement in both operands:
\begin{equation}
\label{orseqmonref}
{\bf A_1 \REF A_2 \wedge A_3 \REF A_4 \ \ \ \implies \ \ \ A_1 \OR A_3 \REF A_2 \OR A_4\ \ \ \text{and}\ \ \  A_1 \SEQ A_3 \REF A_2 \SEQ A_4}
\end{equation}

We first check the monotonicity of the dependency operator $\DEP$ in its left operand. Assume $A_1 \REF A_2$:
\begin{equation*}
\begin{array}{l}
\ \ A_1 \DEP B\\
= \color{magenta}\{\eqref{diff1}\}\\
\ \ [gA_1] \SEQ [gB] \SEQ bA_1 \SEQ [gB] \SEQ bB\\
= \color{magenta}\{\text{assumption properties}\}\\
\ \ [gB] \SEQ [gA_1] \SEQ bA_1 \SEQ [gB] \SEQ bB\\
= \color{magenta}\{\eqref{assumpt}\}\\
\ \ [gB] \SEQ A_1 \SEQ B\\
\REF \color{magenta}\{\SEQ~\text{is monotonic and}~A_1 \REF A_2\}\\
\ \ [gB] \SEQ A_2 \SEQ B\\
= \color{magenta}\{\eqref{diff1},\eqref{assumpt}\}\\
\ \ A_2 \DEP B
\end{array}
\end{equation*}
Hence, we have that the dependency operator is always monotonic in its left operand with respect to refinement:
\begin{equation}
\label{depmonrefleft}
{\bf A_1 \REF A_2 \implies A_1 \DEP B \REF A_2 \DEP B}
\end{equation}

When checking monotonicity for the dependency operator in the right operand, when $A_1 \REF A_2$, we obtain that:
\begin{equation*}
\begin{array}{l}
\ \ B \DEP A_1\\
= \color{magenta}\{\eqref{diff1}\}\\
\ \ [gB] \SEQ [gA_1] \SEQ bB \SEQ [gA_1] \SEQ bA_1\\
= \color{magenta}\{\text{assumption properties},~\eqref{assumpt}\}\\
\ \ [gA_1] \SEQ B \SEQ A_1\\
\REF \color{magenta}\{\SEQ~\text{is monotonic and}~A_1 \REF A_2\}\\
\ \ [gA_1] \SEQ B \SEQ A_2\\
\end{array}
\end{equation*}
\begin{equation*}
\begin{array}{l}
\ \ B \DEP A_2\\
= \color{magenta}\{\eqref{diff1},~\text{assumption properties},~\eqref{assumpt}\}\\
\ \ [gA_2] \SEQ B \SEQ A_2\\
\end{array}
\end{equation*}
Hence, we have that the dependency operator is monotonic in its right operand with respect to refinement conditionally, namely if $gA_2 \implies gA_1$:
\begin{equation}
\label{depmonrefright}
{\bf A_1 \REF A_2 \implies B \DEP A_1 \REF B \DEP A_2  \ \ \ \ \ \ \text{iff} \ \ \ \ \ \ gA_2 \implies gA_1}
\end{equation}

\section{A Train Example}\label{case-study}
\begin{figure}[h]
\centering
\renewcommand{\figurename}{Figure}
\includegraphics[width=0.5\textwidth]{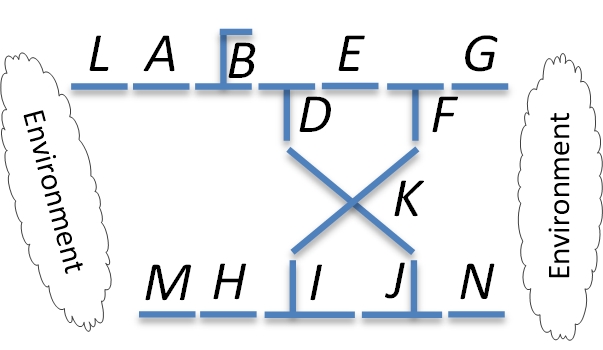}
\caption{The example trajectory \label{CSBC}}
\end{figure}

We now present highlights from a larger case study on token movements on a trajectory. The trajectory is illustrated in Figure~\ref{CSBC}. Tokens can move in any direction and can entry and exit at any of the marginal segments, i.e, $\{L,\ B,\ G,\ M,\ N\}$. For simplicity we assume we only have one token in this paper, that needs to move from $L$ to $N$. We discuss the $\mathcal{TS}$ Action System below, noting that it is inspired by a case study on train movements introduced in~\cite{eventbbook}. Here, we exclude traffic lights, switches and the possibility of one train occupying several segments, but add the possibility of loops. The $\mathcal{TS}$ action system can be thought of as a one-lane road system or communication network.

A token is an element under transportation in the network of slots connected to each other, realistically a packet, a car or anything with a `reverse gear'. A token enters the network from a slot and exits the network from another slot, its destination. The token occupies at most one slot in the network at any given moment. The token is aware of its destination; this is a slot `consuming' the token, like the IP-address in a TCP/IP packet or the end stop of a bus. Each slot is occupied by a token or is `null', i.e., not occupied. Moreover, a non-empty subset of the slots may construct a loop that has a direction of looping, much like a roundabout.

In order for a token to advance from its point of entry towards its point of exit, dependency is crucial. Advancing from a slot to the next is an atomic action. Here the action relies on a token in the current slot but depends on the next slot not to be occupied, i.e. being in slot $A$ and advancing to slot $B$ is a situation where $B$'s slot needs to be non-occupied, written $A \DEP B$. Thus, only if $B$ can provide the service of admitting occupancy to the token, may the token be moved. When a token is advancing in the other direction, the dependency relation is, naturally, inversed: $B \DEP A$. For this, each slot is associated with two actions: $fA$, when a token is moved from slot $A$ and $tA$, with a token is moved to slot $A$. Their corresponding guards are denoted $gfA$, $gtA$ and their corresponding bodies are denoted $bfA$, $btA$, with $f$ for "from" and $t$ for "to".

Consider the set, $SLOTS$ and the string $token$; $SLOTS$ contains the names of the slots. We use the convention that slots are identified with capital letters and tokens with non-capital letters. We initialise these two sets as follows: $SLOTS=\{L, A, B, D, E, G, F, K, I, $ $J, M, H, N\}$ and $token=a$. Here the token occupies the following slots in this order: $L, A, B, D, K, J, X$*, $N$, where $X$ is the loop $I, K, F, E, D, K, J$ and * stands for 0 to $n$ times parsing the loop. Thus, the token may loop any number of times, but shall eventually exit the loop at $J$.

\begin{eqnarray*}
\label{as_case_study}
\mathcal{TS}&=&|[\VAR SLOTS\ :\ {\cal P}(String),\ token\ :\ String, at: String \to String, \\
&& \quad dest: String \to String, neigh: {\cal P}(String \times String) \st\\
&& \quad SLOTS \is \{L,\ A,\ B,\ D,\ E,\ G,\ F,\ K,\ I,\ J,\ M,\ H,\ N\}\SEQ\\
&& \quad token \is a \SEQ at \is \{L \mapsto a, A \mapsto null, ...\}\SEQ\\
&& \quad dest \is \{a \mapsto N\} \SEQ neigh \is \{(L,A),(A,L), ...\}\\
&&\DO \\
&&\ \ \!neigh(L,A) \to fL \DEP tA \\
&&\OR neigh(A,L) \to fA \DEP tL \\
&&\OR neigh(A,B) \to fA \DEP tB  \\
&&\OR neigh(B,D) \to fB \DEP tD \\
&&\OR ...\\
&&\OD]|
\end{eqnarray*}
We have that $fX \DEP tY$ has the following form:
\begin{equation}
\label{fromXtoY}
fX \DEP tY = (at(X)=token \to at(X) \is null) \DEP (at(Y)=null \to at(Y) \is token)
\end{equation}

We notice that all choices are nondeterministic, implying that a token may loop forever instead of exiting. For this, the $\POR$ operator may be used. Thus, we revise the loop exiting action as follows:
$$dest(token)=N \to (fJ \DEP tN) \POR (fJ \DEP tI)$$
Realistically, this means that if the token destination implies an exit, it will exit the loop if the next slot is not occupied.

\section{Conclusions}
In this paper we have initiated a study on a dedicated dependency operator, modeled via the Action Systems formalism. We have studied its commutativity, associativity, and distributivity over other composition operators, as well as its refinement rules. We have illustrated various concepts with several examples.

Action Systems is a state-based formalism, similar in that to Event-B \cite{eventbbook}, Z \cite{Z}, Unity \cite{unity88} (and its MobileUnity \cite{mobileUnity} extension), TLA \cite{Lam94}, etc. One of the most popular such state-based formalisms is Event-B at the moment, justifiably based on the tool support as well as the refinement paradigm it promotes. Action Systems had an attempt at building a theorem prover tool, namely the Refinement Calculator \cite{refcalc}, but that was a bit ahead of its time and was abandoned due to complexity and low interactivity. Action Systems are however highly flexible and versatile, also promoting modularity in a natural manner still not common to Event-B (but out of scope in this paper as well). Being able to model sequentiality, prioritised composition, dependency and reasoning about their properties is a clear advantage that Action Systems provide. Now, the next step is to be able to save all these operators and properties as a theory of actions, for instance via the theory plugin \cite{theoryplugin} in the Rodin platform, and being able to reuse such a theory in various contexts.


{\footnotesize
\begin{thebibliography}{[200]}
\addcontentsline{toc}{section}{\protect\numberline{8}{Bibliography}}

\bibitem{eventbbook} J-R. Abrial, \emph{Modeling in Event-B: System and Software Engineering}. Cambridge University Press, 2010. ISBN-13: 978-0521895569


\bibitem{bbook} J. R. Abrial. {\em The B-Book: Assigning Programs
to Meanings}. Cambridge University Press, 1996. ISBN:0-521-49619-5 


\bibitem{rodin} J-R. Abrial, M. Butler, S. Hallerstede, T. S. Hoang, F. Mehta and L. Voisin. Rodin: An Open Toolset for Modelling and Reasoning in Event-B. In 
{\em International Journal on Software Tools for Technology Transfer (STTT)}, Vol. 12, No. 6, pp 447-466, Springer, 2010. 
\doi {10.1007/s10009-010-0145-y}

\bibitem{parle89} R. J. Back. A Method for Refining Atomicity in
Parallel Algorithms. In {\em E. Odijk, M. Rem, J.-C. Syre (eds), Proceedings of PARLE'89 -- Parallel Architectures and Languages, Vol. 2:
Parallel Languages}, pp. 199-216, 1989.
\doi {10.1007/3-540-51285-3\_42}

\bibitem{rcII} R. J. Back. Refinement Calculus, Part II: Parallel
and Reactive Systems. In {\em J. W. de Bakker, W.-P. de Roever, G.
Rozenberg (eds), Proceedings of Stepwise Refinement of Distributed
Systems: Models, Formalisms, Correctness}, Lecture Notes in
Computer Science, Vol. 430. Springer-Verlag, 1990.
\doi {10.1007/3-540-52559-9\_61}

\bibitem{BaKu83} R. J. Back and R. Kurki-Suonio.
Decentralization of process nets with centralized control. In {\em
Proceedings of the 2nd ACM SIGACT-SIGOPS Symposium on Principles
of Distributed Computing}, pp. 131-142, 1983.
\doi {10.1145/800221.806716}

\bibitem{BaKu88} R. J. Back and R. Kurki-Suonio.
Distributed Cooperation with Action Systems. In {\em ACM
Transactions on Programming Languages and Systems}, Vol. 10, No. 4,
pp. 513-554, 1988.
\doi {10.1145/48022.48023}

\bibitem{procBaSe} R. J. Back and K. Sere. Action Systems with Synchronous
Communication. In {\em E.R. Olderog (ed), Proceedings of PROCOMET'94
-- Programming Concepts, Methods, and Calculi}, pp. 107-126. IFIP
Transactions A-56, North Holland, 1994.

\bibitem{modules} R. J. Back and K. Sere. From Action Systems to Modular Systems.
In {\em Software - Concepts and Tools}, Vol. 17, pp. 26-39,
Springer-Verlag, 1996.
\doi {10.1007/3-540-58555-9\_83}


\bibitem{mpc89} R. J. Back and K. Sere. Stepwise Refinement of
Action Systems. In {\em J. L. A. van de Snepscheut (ed),
Proceedings of MPC'89 -- Mathematics of Program Construction}, pp.
115-138, 1989.
\doi {10.1007/3-540-51305-1\_7}

\bibitem{BSS} M. Butler, E. Sekerinski, and K. Sere. An Action
System Approach to the Steam Boiler Problem. In {\em J.-R. Abrial,
E. B\"{o}rger and H. Langmaack (eds), Formal Methods for
Industrial Applications: Specifying and Programming the Steam
Boiler Control}. Lecture Notes in Computer Science, Vol. 1165,
Springer-Verlag, 1996.
\doi {10.1007/BFb0027234}

\bibitem{refcalc} M. Butler, J. Grundy, T. L\aa ngbacka, R. Ruksenas, and
J. von Wright. The Refinement Calculator: Proof Support for Program
Refinement. In \emph{L. Groves and S. Reeves (eds), Proceedings of
FMP'97 - Formal Methods Pacific}. Discrete Mathematics and
Theoretical Computer Science Series, pp. 40-61, Springer-Verlag,
1997.

\bibitem{unity88} K. M. Chandy and J. Misra. \emph{Parallel Program
Design: A Foundation}. Addison-Wesley, 1988.  ISBN-13: 978-0201058666 

\bibitem{Dijkstra76} E. W. Dijkstra. {\em A Discipline of Programming}.
Prentice Hall International, 1976.  ISBN-13: 978-0132158718 

\bibitem{HKS97} E. Hedman, J. N. Kok, and K. Sere. Coordinating Action Systems.
In {\em D. Garlan and D. Le M\'{e}tayer (eds), Proceedings of
Coordination'97 -- Coordination Languages and Models}, Lecture Notes
in Computer Science, Vol. 1282, pp. 302-319, Springer-Verlag, 1997.
\doi {10.1007/3-540-63383-9\_88}

\bibitem{Hoare69} C.~A.~R.~Hoare. An Axiomatic Basis for Computer
Programming. In {\em Communications of the ACM}, Vol. 12, No. 10,
pp. 576-580, 583, 1969.
\doi {10.1145/363235.363259}

\bibitem{Hoa78} C.A.R. Hoare. Communicating Sequential Processes.
In {\em Communications of the ACM}, Vol. 21, No. 8, pp. 666-677,
1978.

\bibitem{Lam94} L. Lamport. The Temporal Logic of Actions. In {\em
ACM Transactions on Programming Languages and Systems}, Vol. 16, No.
3, pp. 872-923, 1994.
\doi {10.1145/177492.177726}

\bibitem{theoryplugin} I. Maamria, M. Butler, A. Edmunds, and A. Rezazadeh. On an Extensible Rule-based Prover for Event-B. In ABZ2010, Springer, 2010.
\doi {10.1007/978-3-642-11811-1\_40}

\bibitem{MP83} Z. Manna and A. Pnueli. How to cook a temporal proof system
for your pet language. In {\em Proceedings of the Tenth ACM
Conference on Principles of Programming Languages}, pp. 141-154, ACM
New York, 1983.
\doi {10.1145/567067.567082}

\bibitem{Mil80} R. Milner. {\em A Calculus of Communicating Systems}.
Lecture Notes in Computer Science, Vol. 92, Springer-Verlag, 1980. ISBN: 978-3-540-10235-9 

\bibitem{NeSe08} M. Neovius and K. Sere. Formal Modular Modelling of Context-Awareness. In \emph{F. S. de Boer, M. M. Bonsangue and E. Madelain (eds), Formal Methods for Components and Objects, 7th International Symposium,
               {FMCO} 2008}. Lectures Notes in Computer Science, Vol. 5751, pp. 102-118, Springer-Verlag, 2008.
\doi {10.1007/978-3-642-04167-9\_6}

\bibitem{LPthesis05} L. Petre. {\emph Modelling with Action Systems}. TUCS Dissertations No 69, November 2005. ISBN: 951-29-4018-3

\bibitem{mobileUnity} G.-C. Roman and  P. J. McCann. A Notation and Logic for Mobile Computing. In \emph{Formal Methods in System Design}, Vol. 20,
No. 1, pp. 47-68, 2002.
\doi {10.1023/A:1012908529306}

\bibitem{rosese96} M. R\"onkk\"o, E. Sekerinski, and K. Sere.
Control Systems as Action Systems - A Case Study. In \emph{R.
Smedinga, M.P. Spathopoulus, P. Koz\'ak (eds), Proceedings of
WODES'96 -- Workshop on Discret Event Systems}, IEEE Press, pp.
362-367, 1996.

\bibitem{Sekerinski96} E. Sekerinski. Deriving Control Programs by
Weakest Preconditions. \emph{TUCS Technical Reports}, No. 4, 1996.

\bibitem{SS96} E. Sekerinski and K. Sere, A Theory of Prioritizing
Composition. In \emph{The Computer Journal}, Vol. 39, No 8, pp.
701-712. The British Computer Society, Oxford University Press,
1996.
\doi {10.1093/comjnl/39.8.701}

\bibitem{SeWa97} K. Sere and M. Wald\'en. Data Refinement of Remote Procedures.
In {\em M. Abadi and T. Ito (eds.), Proceedings of TACS'97 --
International Symposium on Theoretical Aspects of Computer
Software}, Lecture Notes in Computer Science, Vol. 1281, pp.
267-294, Springer-Verlag, 1997.
\doi {10.1007/PL00003935}


\bibitem{Z} M. Spivey. \emph{The Z Notation: A Reference Manual (Second Edition)}. Prentice Hall International Series in Computer Science, 1992.

\bibitem{WaSe98} M. Wald\'en and K. Sere. Reasoning about Action
Systems using the B-Method. In \emph{Formal Methods in System
Design}, No 13, pp. 5-35. Kluwer Academic Publishers, 1998.
\doi {10.1023/A:1008688421367}


\end{thebibliography}}
\end{document}